\documentclass[12pt]{article}
  \usepackage{amsfonts}
  \usepackage{amsmath}
\usepackage{amssymb}
\usepackage{amscd}
\usepackage[dvips]{graphicx}
\begin{document}

~~
\bigskip
\bigskip
\begin{center}
{\Large {\bf{{{Landau energy levels for twisted N-enlarged Newton-Hooke space-time}}}}}
\end{center}
\bigskip
\bigskip
\bigskip
\begin{center}
{{\large ${\rm {Marcin\;Daszkiewicz}}$}}
\end{center}
\bigskip
\begin{center}
\bigskip

{ ${\rm{Institute\; of\; Theoretical\; Physics}}$}

{ ${\rm{ University\; of\; Wroclaw\; pl.\; Maxa\; Borna\; 9,\;
50-206\; Wroclaw,\; Poland}}$}

{ ${\rm{ e-mail:\; marcin@ift.uni.wroc.pl}}$}

\end{center}
\bigskip
\bigskip
\bigskip
\bigskip
\bigskip
\bigskip
\bigskip
\bigskip
\bigskip
\begin{abstract}
We derive the Landau energy levels for twisted N-enlarged Newton-Hooke space-time, i.e. we find the time-dependent energy
spectrum and the corresponding eigenstates for an electron moving in uniform magnetic as well as in uniform electric external fields.
\end{abstract}
\bigskip
\bigskip
\bigskip
\bigskip
\eject

\section{{{Introduction}}}

The suggestion to use noncommutative coordinates goes back to
Heisenberg and was firstly  formalized by Snyder in \cite{snyder}.
Recently, there were also found formal  arguments based mainly  on
Quantum Gravity \cite{2}, \cite{2a} and String Theory models
\cite{recent}, \cite{string1}, indicating that space-time at Planck
scale  should be noncommutative, i.e. it should  have a quantum
nature. Consequently, there appeared a lot of papers dealing with
noncommutative classical and quantum  mechanics (see e.g.
\cite{mech}-\cite{qm}) as well as with field theoretical models
(see e.g. \cite{prefield}-\cite{fiorewess}) in which  the quantum
space-time is employed.

It is well-known that  a proper modification of the Poincare and
Galilei Hopf algebras can be realized in the framework of Quantum
Groups \cite{qg1}, \cite{qg3}. Hence, in accordance with the
Hopf-algebraic classification  of all deformations of relativistic
and nonrelativistic symmetries (see \cite{class1}, \cite{class2}),
one can distinguish three
types of quantum spaces \cite{class1}, \cite{class2} (for details see also \cite{nnh}):\\
\\
{ \bf 1)} Canonical ($\theta^{\mu\nu}$-deformed) type of quantum space \cite{oeckl}-\cite{dasz1}
\begin{equation}
[\;{ x}_{\mu},{ x}_{\nu}\;] = i\theta_{\mu\nu}\;, \label{noncomm}
\end{equation}
\\
{ \bf 2)} Lie-algebraic modification of classical space-time \cite{dasz1}-\cite{lie1}
\begin{equation}
[\;{ x}_{\mu},{ x}_{\nu}\;] = i\theta_{\mu\nu}^{\rho}{ x}_{\rho}\;,
\label{noncomm1}
\end{equation}
and\\
\\
{ \bf 3)} Quadratic deformation of Minkowski and Galilei  spaces \cite{dasz1}, \cite{lie1}-\cite{paolo}
\begin{equation}
[\;{ x}_{\mu},{ x}_{\nu}\;] = i\theta_{\mu\nu}^{\rho\tau}{
x}_{\rho}{ x}_{\tau}\;, \label{noncomm2}
\end{equation}
with coefficients $\theta_{\mu\nu}$, $\theta_{\mu\nu}^{\rho}$ and  $\theta_{\mu\nu}^{\rho\tau}$ being constants.\\
\\
Moreover, it has been demonstrated in \cite{nnh}, that in the case of
so-called N-enlarged Newton-Hooke Hopf algebras
$\,{\mathcal U}^{(N)}_0({ NH}_{\pm})$ the twist deformation
provides the new  space-time noncommutativity of the
form\footnote{$x_0 = ct$.},\footnote{ The discussed space-times have been  defined as the quantum
representation spaces, so-called Hopf modules (see e.g. \cite{oeckl}, \cite{chi}), for quantum N-enlarged
Newton-Hooke Hopf algebras.}
\begin{equation}
{ \bf 4)}\;\;\;\;\;\;\;\;\;[\;t,{ x}_{i}\;] = 0\;\;\;,\;\;\; [\;{ x}_{i},{ x}_{j}\;] = 
if_{\pm}\left(\frac{t}{\tau}\right)\theta_{ij}(x)
\;, \label{nhspace}
\end{equation}
with time-dependent  functions
$$f_+\left(\frac{t}{\tau}\right) =
f\left(\sinh\left(\frac{t}{\tau}\right),\cosh\left(\frac{t}{\tau}\right)\right)\;\;\;,\;\;\;
f_-\left(\frac{t}{\tau}\right) =
f\left(\sin\left(\frac{t}{\tau}\right),\cos\left(\frac{t}{\tau}\right)\right)\;,$$
$\theta_{ij}(x) \sim \theta_{ij} = {\rm const}$ or
$\theta_{ij}(x) \sim \theta_{ij}^{k}x_k$ and  $\tau$ denoting the time scale parameter
 -  the cosmological constant. Besides, it should be  noted that the mentioned above quantum spaces {\bf 1)}, { \bf 2)} and { \bf 3)}
 can be obtained  by the proper contraction limit  of the commutation relations { \bf 4)}\footnote{Such a result indicates that the twisted N-enlarged Newton-Hooke Hopf algebra plays a role of the most general type of quantum group deformation at nonrelativistic level.}.

 As it was mentioned above, recently, there has been discussed the impact of different  kinds of
quantum spaces on the dynamical structure of physical systems (see e.g. \cite{mech}-\cite{field}). However, the especially interesting results have been obtained in the series of papers
\cite{hallcan}-\cite{hallcanf} concerning the Hall effect for canonically deformed space-time (\ref{noncomm}). Particularly, there has been found the $\theta$-dependent
energy spectrum of an electron moving in uniform magnetic as well as in uniform electric field. Besides, it was demonstrated that in the case of noncommutative system  one obtains the standard (commutative) Hall conductivity in terms of an  effective magnetic field $B_{\rm {eff}}(\theta)$.

In this article we derive the Landau energy levels for twisted N-enlarged Newton-Hooke space-time (\ref{spaces}). Preciously, we find the time-dependent energy spectrum and the corresponding
time-dependent eigenfunctions for an electron moving in uniform electric as well as magnetic field.  Of course, for special choice of the noncommutativity function
$f(t)$ (see formula (\ref{spaces})) i.e. for $f(t) = \theta$, we rediscovery  partially the results of articles \cite{hallcan}-\cite{hallcanf}.

The paper is organized as follows. In Sect. 2 we recall basic facts
concerning the twisted N-enlarged Newton-Hooke   space-times
provided in article \cite{nnh}. The third section is devoted to the calculation of Landau energy levels for commutative (classical) space, while the Landau spectrum for  twisted N-enlarged Newton-Hooke   space-times is derived in Sect. 4. The
final remarks are presented in the last section.

\section{Twisted N-enlarged Newton-Hooke space-times}

In this section we recall the basic facts associated with the twisted N-enlarged Newton-Hooke Hopf algebra $\;{\cal U}^{(N)}_{\alpha}(NH_{\pm})$ and with the
corresponding quantum space-times \cite{nnh}.  Firstly, it should be noted that in accordance with Drinfeld  twist procedure the algebraic sector of twisted
Hopf structure $\;{\cal U}^{(N)}_{\alpha}(NH_{\pm})$ remains
undeformed, i.e. it takes the form
 \begin{eqnarray}
&&\left[\, M_{ij},M_{kl}\,\right] =i\left( \delta
_{il}\,M_{jk}-\delta _{jl}\,M_{ik}+\delta _{jk}M_{il}-\delta
_{ik}M_{jl}\right)\;\; \;, \;\;\; \left[\, H,M_{ij}\,\right] =0
 \;,  \label{q1} \\
&~~&  \cr &&\left[\, M_{ij},G_k^{(n)}\,\right] =i\left( \delta
_{jk}\,G_i^{(n)}-\delta _{ik}\,G_j^{(n)}\right)\;\; \;, \;\;\;\left[
\,G_i^{(n)},G_j^{(m)}\,\right] =0 \;,\label{q2}
\\
&~~&  \cr &&\left[ \,G_i^{(k)},H\,\right] =-ikG_i^{(k-1)}\;\; \;, \;\;\; \left[\, H,G_i^{(0)}\,\right] =\pm \frac{i}{\tau}G_i^{(1)}\;\;\;;\;\;\;k>1
\;,\label{q3}
\end{eqnarray}
where $\tau$, $M_{ij}$, $H$, $G_i^{(0)} (=P_i)$, $G_i^{(1)} (=K_i)$ and $G_i^{(n)} (n>1)$ can be identified with cosmological time parameter, rotation, time translation, momentum, boost and accelerations  operators respectively. Besides, the coproducts and antipodes of considered algebra are given by\footnote{$\Delta_0(a) = a\otimes 1 +1\otimes a$, $S_{0}(a) =-a$.}
\begin{equation}
 \Delta _{\alpha }(a) = \mathcal{F}_{\alpha }\circ
\,\Delta _{0}(a)\,\circ \mathcal{F}_{\alpha }^{-1}\;\;\;,\;\;\;
S_{\alpha}(a) =u_{\alpha }\,S_{0}(a)\,u^{-1}_{\alpha }\;,\label{fs}
\end{equation}
with $u_{\alpha }=\sum f_{(1)}S_0(f_{(2)})$ (we use Sweedler's notation
$\mathcal{F}_{\alpha }=\sum f_{(1)}\otimes f_{(2)}$) and with the twist factor
$\mathcal{F}_{\alpha } \in {\cal U}^{(N)}_{\alpha}(NH_{\pm}) \otimes
{\cal U}^{(N)}_{\alpha}(NH_{\pm})$
satisfying  the classical cocycle condition
\begin{equation}
{\mathcal F}_{{\alpha }12} \cdot(\Delta_{0} \otimes 1) ~{\cal
F}_{\alpha } = {\mathcal F}_{{\alpha }23} \cdot(1\otimes \Delta_{0})
~{\mathcal F}_{{\alpha }}\;, \label{cocyclef}
\end{equation}
and the normalization condition
\begin{equation}
(\epsilon \otimes 1)~{\cal F}_{{\alpha }} = (1 \otimes
\epsilon)~{\cal F}_{{\alpha }} = 1\;, \label{normalizationhh}
\end{equation}
such that ${\cal F}_{{\alpha }12} = {\cal F}_{{\alpha }}\otimes 1$ and
${\cal F}_{{\alpha }23} = 1 \otimes {\cal F}_{{\alpha }}$.

The corresponding quantum space-times are defined as the representation spaces (Hopf modules) for N-enlarged Newton-Hooke Hopf algebra
\;${\cal U}_{\alpha}^{(N)}(NH_{\pm})$. Generally, they are equipped with two spatial directions
commuting to classical time, i.e. they take  the form 
\begin{equation}
[\;t,\hat{x}_{i}\;] =[\;\hat{x}_{1},\hat{x}_{3}\;] = [\;\hat{x}_{2},\hat{x}_{3}\;] =
0\;\;\;,\;\;\; [\;\hat{x}_{1},\hat{x}_{2}\;] =
if({t})\;\;;\;\;i=1,2,3
\;. \label{spaces}
\end{equation}
 However, it should be noted 
that this type of noncommutativity  has  been  constructed explicitly  only in the case of 6-enlarged Newton-Hooke Hopf algebra, with 
\cite{nnh}\footnote{$\kappa_a = \alpha$ $(a=1,...,36)$ denote the deformation parameters.}
\begin{eqnarray}
f({t})&=&f_{\kappa_1}({t}) =
f_{\pm,\kappa_1}\left(\frac{t}{\tau}\right) = \kappa_1\,C_{\pm}^2
\left(\frac{t}{\tau}\right)\;, \nonumber
\end{eqnarray}
\begin{eqnarray}
f({t})&=&f_{\kappa_2}({t}) =
f_{\pm,\kappa_2}\left(\frac{t}{\tau}\right) =\kappa_2\tau\, C_{\pm}
\left(\frac{t}{\tau}\right)S_{\pm} \left(\frac{t}{\tau}\right) \;,
\nonumber\\
&~~&~~~~~~~~~~~~~~~~~~~~~~~~~~~~~~~~~ \nonumber\\
&~~&~~~~~~~~~~~~~~~~~~~~~~~~~~~~~~~~~\cdot \nonumber\\
&~~&~~~~~~~~~~~~~~~~~~~~~~~~~~~~~~~~~\cdot \label{w2}\\
&~~&~~~~~~~~~~~~~~~~~~~~~~~~~~~~~~~~~\cdot \nonumber\\
&~~&~~~~~~~~~~~~~~~~~~~~~~~~~~~~~~~~~ \nonumber\\
f({t})&=&
f_{\kappa_{35}}\left(\frac{t}{\tau}\right) = 86400\kappa_{35}\,\tau^{11}
\left(\pm C_{\pm} \left(\frac{t}{\tau}\right)  \mp \frac{1}{24}\left(\frac{t}{\tau}\right)^4 - \frac{1}{2}
\left(\frac{t}{\tau}\right)^2 \mp 1\right) \,\times \nonumber\\
&~~&~~~~~~~~~~~~~~~~\times~\;\left(S_{\pm} \left(\frac{t}{\tau}\right)  \mp \frac{1}{6}\left(\frac{t}{\tau}\right)^3 - \frac{t}{\tau}\right)\;,
\nonumber\\
f({t})&=&
f_{\kappa_{36}}\left(\frac{t}{\tau}\right) =
518400\kappa_{36}\,\tau^{12}\left(\pm C_{\pm} \left(\frac{t}{\tau}\right)  \mp \frac{1}{24}\left(\frac{t}{\tau}\right)^4 - \frac{1}{2}
\left(\frac{t}{\tau}\right)^2 \mp 1\right)^2\;, \nonumber
\end{eqnarray}
and
$$C_{+/-} \left(\frac{t}{\tau}\right) = \cosh/\cos \left(\frac{t}{\tau}\right)\;\;\;{\rm and}\;\;\;
S_{+/-} \left(\frac{t}{\tau}\right) = \sinh/\sin
\left(\frac{t}{\tau}\right) \;.$$
Besides, one can easily check that in $\tau$ approaching infinity limit the above quantum spaces reproduce the canonical (\ref{noncomm}),
Lie-algebraic (\ref{noncomm1}) and quadratic (\ref{noncomm2})  type of
space-time noncommutativity, i.e. for $\tau \to \infty$ we get
\begin{eqnarray}
f_{\kappa_1}({t}) &=& \kappa_1\;,\nonumber\\
f_{\kappa_2}({t}) &=& \kappa_2\,t\;,\nonumber\\
&\cdot& \nonumber\\
&\cdot& \label{qqw2}\\
&\cdot& \nonumber\\
f_{\kappa_{35}}({t}) &=& \kappa_{35}\,t^{11}\;, \nonumber\\
f_{\kappa_{36}}({t}) &=& \kappa_{36}\,t^{12}\;. \nonumber
\end{eqnarray}
Of course, for all deformation parameters $\alpha = \kappa_a$  running to zero the above deformations disappear.

\section{{{Landau energy levels for commutative space-time}}}

In this section we turn to  the derivation of Landau energy levels for commutative (classical) space. Firstly, it should be noted, that in such a case
the algebra of position and momentum operators takes the form
\begin{equation}
[\;x_{i},x_{j}\;] = 0 =[\;p_{i},p_{j}\;]\;\;\;,\;\;\; [\;x_{i},p_{j}\;]
={i\hbar}\delta_{ij}\;. \label{classpoisson}
\end{equation}
Further, we define the following Hamiltonian function
\begin{eqnarray}
H=H(\bar{p}, \bar{x}) = {1\over 2m}\left({\bar p}+{e\over c}{\bar A}(\bar{x})\right)^2
-e\phi(\bar{x})\;, \label{os1}
\end{eqnarray}
which describes an electron moving in the $(x_1,x_2)$-plane in the uniform external electric field ${\bar E} = -{\rm grad} \phi$ and in the uniform perpendicular to plane
 magnetic field ${\bar B} = {\rm rot} {\bar A}$. Besides, in our considerations we adopt the so-called symmetric gauge
\begin{eqnarray}
{\bar A}(\bar{x})=\left[-{B\over 2}x_2,{B\over 2}x_1\right]\;, \label{os2}
\end{eqnarray}
as well as we take
\begin{eqnarray}
\phi(\bar{x}) =-Ex_1\;. \label{os3}
\end{eqnarray}
Consequently, we get
\begin{eqnarray}
H(\bar{p}, \bar{x}) =
\frac{1}{2m}\left[
\left( p_1 -\frac{eB}{2c} x_2 \right)^2 +
\left( p_2 +\frac{eB}{2c} x_1 \right)^2 \right] +eEx_1\;. \label{os4}
\end{eqnarray}

Let us now turn to the eigenvalue problem for hamiltonian function (\ref{os4}) encoded by
\begin{eqnarray}
{H}(\bar{p}, \bar{x}) \psi(\bar{x}) ={\mathcal{E}}\psi(\bar{x})\;. \label{os5}
\end{eqnarray}
In order to solve the above equation we perform the following change of variables
\begin{eqnarray}
x={x}_1+i{x}_2\;\;\;,\;\;\; {p} =\frac{1}{2} ({p}_1
-i{p}_2)\;, \label{os6}
\end{eqnarray}
as well as we introduce two families of creation/annihilation operators
\begin{eqnarray}
a^\dag =-2i{p}^*+{eB\over 2c}x+\lambda \;\;\;,\;\;\;
a =2i{p}+{eB\over 2c}x^*+\lambda\;, \label{os7}
\end{eqnarray}
and
\begin{eqnarray}
b=2ip-{eB\over 2c}x^*\;\;\;,\;\;\;
b^{\dag}=-2i{p}^*-{eB\over 2c}x\;\;\;;\;\;\;(\alpha + i\beta)^* = \alpha -i\beta\;, \label{os8}
\end{eqnarray}
with parameter $\lambda = {mcE\over B}$.  One can easily check that these two sets commute with each other and satisfy the following commutation relations
\begin{eqnarray}
[\;a, a^{\dag}\;]=2m\hbar\omega\;\;\;,\;\;\;[\;b^{\dag}, b\;]=2m\hbar\omega
\;, \label{os8}
\end{eqnarray}
with $\omega={eB\over mc}$ denoting  cyclotron frequency. Besides, we observe that the Hamiltonian ${H}(\bar{p}, \bar{x})$ can be written as
\begin{eqnarray}
{H}(a,b)= {1\over 4m}\left(a^{\dag}a+aa^{\dag}\right)-
{\lambda\over 2m}\left(b^{\dag}+b\right)-{\lambda^2\over 2m}
\;. \label{os9}
\end{eqnarray}
In order to find the eigenvalues ${\mathcal{E}}$ and eigenfunctions $\psi(\bar{x})$ we separate the operator (\ref{os9}) into two mutually commuting  parts
\begin{eqnarray}
{H}(a,b)= H(a) + H(b)
\;, \label{os10}
\end{eqnarray}
where $H(a)$ denotes the harmonic oscillator part
\begin{eqnarray}
{H}(a)={1\over 4m}\left(a^{\dag}a+aa^{\dag}\right)
\;, \label{os11}
\end{eqnarray}
while $H(b)$ - the part linear in $b$ and $b^\dag$ operators, given by
\begin{eqnarray}
{H}(b)={\lambda\over 2m}\left(b^{\dag}+b\right) + \frac{\lambda^2}{2m}
\;. \label{os12}
\end{eqnarray}
In the case of the harmonic oscillator part
\begin{eqnarray}
{H}(a)\psi_n={\mathcal{E}}_n\psi_n
\;, \label{os13}
\end{eqnarray}
one can proceed in the standard way and gets the following discrete spectrum
\begin{eqnarray}
\phi_n= \frac{1}{\sqrt{(2m\hbar\omega)^n n!}}(a^{\dag})^n|0>\;\;\;,\;\;\;
{\mathcal{E}}_n={\hbar\omega\over 2} (2n+1)\;\;;\;\; n=0,1,2 \ldots
\;, \label{os14}
\end{eqnarray}
with vacuum state $|0>$ such that $a|0> =0$. For the eigenvalue equation
\begin{eqnarray}
{H}(b)\psi={\mathcal{E}}\psi
\;, \label{os15}
\end{eqnarray}
the situation seems to be more complicated. However, by simple calculation we find the following continuous  spectrum
\begin{eqnarray}
\psi_{\alpha}=\exp{i\left(\alpha x_2+{m\omega\over 2\hbar}x_1x_2\right)}\;\;\;,\;\;\;
{\mathcal{E}}_{\alpha}={\hbar\lambda\over m}\alpha+{\lambda^2\over 2m}
\;, \label{os16}
\end{eqnarray}
with real parameter $\alpha$. Consequently, the eigenfunctions and the energy spectrum of the whole Hamiltonian $H(a,b)$ are given by
\begin{eqnarray}
\psi_{(n,\alpha)}(\bar{x})=\psi_n\otimes \psi_{\alpha}\;\;\;,\;\;\;
{\mathcal{E}}_{(n,\alpha)}={\hbar\omega\over 2}(2n+1)-
{\hbar\lambda\over m}\alpha-{\lambda^2\over 2m}
\;, \label{os17}
\end{eqnarray}
where $n=0,1,2 \ldots$, $\alpha\in  \mathbb{R}$ and where symbol $\otimes$ denotes the direct product of two wave functions. The formula
(\ref{os17}) defines so-called Landau energy levels for commutative nonrelativistic space-time (\ref{classpoisson}).

\section{Landau energy levels for twisted N-enlarged Ne-wton-Hooke space-time}

Let us now turn to the main aim of our investigations - to the derivation of Landau energy levels for quantum space-times (\ref{spaces}).
In first step of our calculation we extend the described in second section spaces to the whole algebra of momentum and position operators as follows
\begin{eqnarray}
&&[\;\hat{ x}_{1},\hat{ x}_{2}\;] = if_{\kappa_a}({t})\;\;\;,\;\;\;
 [\;\hat{ p}_{i},\hat{ p}_{j}\;] =0\;\;\;,\;\;\;[\;\hat{ x}_{i},\hat{ p}_{j}\;] = {i\hbar}\delta_{ij}\;. \label{phasespaces}
\end{eqnarray}
One can check that relations (\ref{phasespaces}) satisfy the Jacobi identity and for deformation parameters
$\kappa_a$ approaching zero become classical. \\
Next, by analogy to the commutative case (see formulas (\ref{os1}), (\ref{os2})) we define the following Hamiltonian operator\footnote{The choice of the Hamiltonian
(\ref{grom1}) (similar to the classical operator (\ref{os1})) permits to investigate the (direct) impact of noncommutative space-time
on the dynamical structure of considered system. As we shall see for a moment such a defined Hamiltonian in terms of the commutative position operators becomes time-dependent (see formula (\ref{grom2})). This situation is interpreted by the author as the generating by space-time noncommutativity (\ref{spaces}) of additional (time-dependent) interaction of particle with some external source.},\footnote{We define the model only
in noncommutative $({\hat x}_1,{\hat x}_2)$-plane, i.e. $\bar{\hat{p}} = ({\hat{p}}_1,{\hat{p}}_2)$ and $\bar{\hat{x}} = ({\hat{x}}_1,{\hat{x}}_2)$.}
\begin{eqnarray}
\hat{H} = \hat{H}(\bar{\hat{p}},\bar{\hat{x}}) = {1\over 2m}\left(\bar{\hat{p}}+{e\over c}\bar{\hat{A}}(\bar{\hat{x}})\right)^2
-e{\hat \phi}(\bar{\hat{x}})
\;.\label{grom1}
\end{eqnarray}
In order to analyze the above system we represent the
noncommutative operators $({\hat x}_i, {\hat p}_i)$ by classical
ones $({ x}_i, { p}_i)$ as  (see e.g.
\cite{giri})
\begin{equation}
{\hat x}_{1} = { x}_{1} - \frac{f_{\kappa_a}(t)}{2\hbar}
p_2\;\;\;,\;\;\;{\hat x}_{2} = { x}_{2} +\frac{f_{\kappa_a}(t)}{2\hbar}
p_1 \;\;\;,\;\;\; {\hat p}_{i}=
p_i\;.\label{rep}
\end{equation}
Then, the Hamiltonian (\ref{grom1}) in the symmetric gauge (\ref{os2}) takes the form
\begin{eqnarray}
\hat{H}(t) &=&
\frac{1}{2m}\left[
\left( (1-\alpha_{\kappa_a}(t) ){p}_1 -\frac{eB}{2c} {x_2} \right)^2 +
\left( (1-\alpha_{\kappa_a}(t) ){p}_2 +\frac{eB}{2c} {x_1} \right)^2 \right] + \nonumber \\
&+&eE\left({x_1}-{f_{\kappa_a}(t)\over 2\hbar}{p}_2\right) = \hat{H}(\bar{{p}},\bar{{x}},t)
\;,\label{grom2}
\end{eqnarray}
with  function $\alpha_{\kappa_a}(t) = \frac{ef_{\kappa_a}(t) B}{4\hbar c}$.

In order to solve the eigenvalue problem
\begin{eqnarray}
\hat{H}(t)\psi={\mathcal{E}}\psi
\;, \label{grom3}
\end{eqnarray}
we introduce (similar to the commutative case) two sets of time-dependent operators
\begin{eqnarray}
a^\dag(t) =-2i{\bar{p}}^*(t)+{eB\over 2c}x+\lambda_{-}(t) \;\;\;,\;\;\;
a(t) =2i{\bar{p}}(t)+{eB\over 2c}x^*+\lambda_{-}(t)\;,
\;, \label{grom4}
\end{eqnarray}
as well as
\begin{eqnarray}
b(t)=2i\bar{p}(t)-{eB\over 2c}x^*\;\;\;,\;\;\;
b^{\dag}(t)=-2i{\bar{p}}^*(t)-{eB\over 2c}x
\;, \label{grom5}
\end{eqnarray}
where $\bar{p}(t) = \beta(t)p  = (1 - \alpha_{\kappa_a}(t))p$ and where $\lambda_{-}(t)$ denotes the real and arbitrary function  which will be fixed for a moment. One can check that both
families of operators $(a(t),a^\dag(t))$ and $(b(t),b^\dag(t))$ commute each other and satisfy the following commutation relations\footnote{We perform our  calculations for such times that $\omega(t) > 0$. Such a situation appears for "almost" all times  for  $0 < \kappa_a <<1$.}
\begin{eqnarray}
[\;{ a}(t) ,{a}^{\dag}(t)\;]=2m\hbar{{\omega}(t)}\;\;\;,\;\;\;
[\;{ b}^{\dag}(t) ,{b}(t)\;]=2m\hbar{\omega (t)}\;, \label{grom6}
\end{eqnarray}
with $\omega (t) = \beta(t)\omega$. Moreover, it is easy to see that Hamiltonian $\hat{H}(t)$ can be written as
\begin{eqnarray}
\hat{H}(a(t),b(t))= {1\over 4m}\left(a^{\dag}(t)a(t)+a(t)a^{\dag}(t)\right)-
{\lambda_{+}(t)\over 2m}\left(b^{\dag}(t)+b(t)\right)-{\lambda_{-}^2(t)\over 2m}
\;, \label{grom7}
\end{eqnarray}
where  functions $\lambda_{\pm}(t)$ are fixed to be
\begin{eqnarray}
\lambda_{\pm}(t) = \lambda \pm \frac{emEf_{\kappa}(t)}{4\beta(t) \hbar}
\;. \label{grom8}
\end{eqnarray}
The solution of eigenvalue problem (\ref{grom3}) can be found by direct calculation;
it looks as follows\footnote{Similar to the commutative case we use the formula (\ref{grom7}) and $p_i = -i\hbar \partial_i$.}
\begin{eqnarray}
\psi_{(n,\alpha,\kappa_a)}(t) &=&
{1\over \sqrt{(2m\hbar{\omega}(t))^n n!}}
\exp{i\left(\alpha x_2+{m{\omega(t)}\over 2\hbar\beta^2(t)}x_1x_2\right)}({a}^{\dag}(t))^n|0>\;, \nonumber \\
{\mathcal{E}}_{(n,\alpha,\kappa_a)}(t)&=&
{\hbar{\omega(t)}\over 2}(2n+1)-
{\hbar\beta(t)\lambda_+(t)\over m}\alpha-{m\over 2}\lambda_-^2(t)
\;, \label{grom9}
\end{eqnarray}
with $n=0,1,2 \ldots$ and $\alpha\in  \mathbb{R}$. The formula (\ref{grom9}) defines the Landau energy levels for
twisted N-enlarged Newton-Hooke space-time (\ref{spaces}). Besides, it should  be noted, that for $f_{\kappa_a}(t) = \theta$ we recover the energy spectrum for canonical deformation derived in
paper \cite{hallcan}, while for all parameters $\kappa_a$ approaching zero the above results become the same as in commutative case, i.e. we get
 the formula (\ref{os17}).

\section{Final remarks}

In this article we derive the Landau energy levels for twisted N-enlarged Newton-Hooke space-time. Preciously, we find the time-dependent energy spectrum
for an electron moving in uniform magnetic as well as in uniform electric (external) fields.  It should be also mentioned that the presented investigation
has been performed for the quite general (constructed explicitly) type of space-time noncommutativity at nonrelativistic level.

\section*{Acknowledgments}
 This paper has been financially  supported  by Polish
NCN grant No 2011/01/B/ST2/03354.

\end{document}